\documentclass[12pt]{article}

\usepackage{amsmath}
\usepackage{amsfonts}
\usepackage{graphicx}
\usepackage{caption}
\usepackage{subcaption}

\usepackage{authblk}
\usepackage{amssymb}

\usepackage{epstopdf}
\usepackage{epsfig}
\usepackage{hyperref}

\usepackage{color}

\date{}

\begin{document}

\title{Axial quasi-normal modes of neutron stars in $R^2$ gravity}

\author[1]{Jose Luis Bl\'azquez-Salcedo \thanks{\href{mailto:jose.blazquez.salcedo@uni-oldenburg.de}{jose.blazquez.salcedo@uni-oldenburg.de}}}

\author[2,3]{Daniela D. Doneva
\thanks{\href{mailto:daniela.doneva@uni-tuebingen.de }{daniela.doneva@uni-tuebingen.de }}}

\author[1]{Jutta Kunz \thanks{\href{mailto:jutta.kunz@uni-oldenburg.de}{jutta.kunz@uni-oldenburg.de}}} 

\author[4]{Kalin V. Staykov \thanks{\href{mailto:kstaykov@phys.uni-sofia.bg}{kstaykov@phys.uni-sofia.bg}}}

\author[2,4,5]{Stoytcho S. Yazadjiev \thanks{\href{mailto:yazad@phys.uni-sofia.bg}{yazad@phys.uni-sofia.bg}}}

\affil[1]{Institut f\"ur  Physik, Universit\"at Oldenburg, Postfach 2503,
D-26111 Oldenburg, Germany}
\affil[2]{Theoretical Astrophysics, Eberhard Karls University of T\"ubingen, T\"ubingen 72076, Germany}
\affil[3]{INRNE - Bulgarian Academy of Sciences, 1784  Sofia, Bulgaria}
\affil[4]{Department of Theoretical Physics, Faculty of Physics, Sofia University, Sofia 1164, Bulgaria}
\affil[5]{Institute of Mathematics and Informatics, Bulgarian Academy of Sciences, Acad. G. Bonchev Street 8, Sofia 1113, Bulgaria}

%\date{\today}

\maketitle

\begin{abstract}
In the present paper the axial quasi-normal modes of neutron stars in $f(R)$ 
gravity are examined using a large set of equations of state. The numerical 
calculations are made using two different approaches -- performing time 
evolution of the perturbation equations and solving the time-independent 
representation of the equations as a boundary value problem. According to the 
results the mode frequencies and the damping times decrease with the increase 
of the free parameter of the theory in comparison to the pure general 
relativistic case. While the frequencies deviate significantly from Einstein's 
theory for all realistic neutron star masses (say above $1M_\odot$), 
the damping times reach non-negligible differences only for the more massive 
models. We have constructed as well universal (equation of state independent) 
gravitational wave asteroseismology relations involving the frequencies and 
the damping times. It turns out that the equation of state independence is 
preserved using the same normalization as in pure general relativity and the 
qualitative differences of the phenomenological relations with respect to 
Einstein's theory of gravity can be large for large values of the free 
parameter in $f(R)$ gravity. 
\end{abstract}
%\pacs{}

\section{Introduction}

Until recently the electromagnetic channel was the only way to probe the strong field regime of gravity. This changed with the direct detection of gravitational waves from merging binary black holes 
\cite{Abbott:2016blz,Abbott:2016nmj,Abbott:2017oio,Abbott:2017vtc,Abbott:2017gyy} 
and binary neutron stars \cite{TheLIGOScientific:2017qsa}. 
A very interesting fact is that in the latter case the electromagnetic 
counterpart was observed as well \cite{GBM:2017lvd}, 
which sets the beginning of the so-called multi-messenger 
gravitational wave astronomy. 
It is therefore most important to study the different sources and mechanisms 
for the emission of gravitational waves. A rather promising avenue here 
is represented by the study of quasi-normal modes (QNMs) of neutron stars. 
Even though such QNMs have not been detected yet, it is expected 
that the advance of the instruments will allow to observe them in the future. 
A drawback of the attempts to test the strong field regime of gravity 
with neutron star QNMs are the uncertainties in the nuclear matter 
equation of state (EOS). However, the observational constraints 
on the EOS are rapidly improving with time 
\cite{Demorest:2010bx,Antoniadis:2013pzd,Lattimer:2013hma,Ozel:2016oaf,Most:2018hfd}. 
Moreover, the QNM spectrum of neutron stars is much richer 
than the black hole one, which offers us different possibilities 
to test deviations from pure general relativity (GR)
(see e.g., \cite{Berti:2015itd,Berti:2018cxi,Berti:2018vdi}).

Neutron star oscillations have been studied for decades, 
and comprehensive classifications have been made 
\cite{Kokkotas:1999bd, Nollert:1999ji, Ferrari:2007dd}. 
In the present paper we will concentrate on the axial modes of neutron stars. 
(Since we will be considering nonrotating models the perturbations can be 
separated into purely axial and polar perturbations.) 
The axial modes are rapidly damped spacetime (or curvature) modes 
that have no analog in Newtonian gravity and are a pure manifestation 
of the tensorial character of GR and its generalizations 
similar to the QNMs of black holes 
\cite{Kokkotas:2003mh,Kokkotas:1994an,Andersson:1995wu}. 
Most of the previous studies of axial QNMs or neutron stars
have been made in pure GR though. 
Exceptions are the studies in massless scalar-tensor theories 
\cite{Sotani:2005qx}, 
tensor-vector-scalar theory \cite{Sotani:2009nm},
Einstein-Gauss-Bonnet-dilaton gravity \cite{Blazquez-Salcedo:2015ets} 
and Horndeski gravity \cite{Blazquez-Salcedo:2018tyn}. 

In the present paper we will concentrate on a class of alternative theories 
of gravity that attracted considerable interest recently in connection 
with the dark energy puzzle, and which can also produce non-negligible 
deviations from GR in the strong field regime, namely the $f(R)$ theories 
of gravity \cite{Sotiriou:2008rp,DeFelice:2010aj,Capozziello:2011et}. 
More precisely, we will focus on the so-called $R^2$ gravity 
having a Lagrangian of the form $f(R) = R + aR^2$, where $a$ is a parameter, 
since the $R^2$ term is supposed to give the dominant contribution 
for strong fields. Realistic non-perturbative neutron star models 
were constructed in $R^2$ gravity in 
\cite{Yazadjiev:2014cza,Staykov:2014mwa,Yazadjiev:2015zia,Astashenok:2017dpo} 
both in the static and rotating cases, 
and the results show that the deviations from GR can be significant, 
even though comparable with the equation of state uncertainty. 

In the present paper we will perturb the non-rotating background models 
obtained in \cite{Yazadjiev:2014cza}.
First we will solve the time-dependent perturbation equation
to obtain the time evolution for the gravitational wave signal,
extracting the oscillation frequency and the damping rate
for some selected neutron star models and a realistic EOS.
Then we will turn to the time-independent boundary value problem,
solving for the QNMs for a large number of models and EOS.
This will then allow us to address universal (EOS independent) relations 
for the neutron star models
(see e.g.~the recent reviews \cite{Yagi:2016bkt,Doneva:2017}).
In particular, we will investigate how
GR universal relations for the rescaled frequencies and damping times 
\cite{Andersson:1996pn,Andersson:1997rn,Kokkotas:1999mn,Benhar:2004xg,BlazquezSalcedo:2012pd,Blazquez-Salcedo:2013jka}
generalize to $f(R)$ theory.

The outline of the paper is as follows:
In Section II the basic equations for the axial perturbations in 
$f(R)$ gravity and some notes on the numerical methods are presented. 
In Section III the numerical results are presented and discussed. 
The paper ends with a conclusion.

\section{Basics}

\subsection{Background solution}

In this paper we study the axial perturbations of neutron stars 
in $f(R)$ gravity with Lagrangian $f(R) = R + a R^2$, 
also known as $R$-squared gravity. 
For mathematical simplicity we will not work directly with 
the $f(R)$ gravity action and field equations, 
but instead we will exploit the mathematical equivalence 
between the $f(R)$ theories and a certain class of 
scalar-tensor theories (STT). 
In addition, when working with the STT representation of the $f(R)$ theories 
we will not employ the physical Jordan frame 
but instead the more convenient Einstein frame. 
Transformations to the relevant quantities in the physical Jordan frame 
will be made whenever needed. 
More details on this problem, 
including a detailed discussion of the transition from $f(R)$ to STT 
and the connection between the Jordan and Einstein frames 
can, e.g., be found in 
\cite{Yazadjiev:2014cza, Staykov:2014mwa, Yazadjiev:2015zia}.

The general form of the Einstein frame action in STT 
can be written in the following form
\begin{eqnarray} \label{eq:action}
S = \frac{1}{16\pi G}\int d^4x\sqrt{-g}\left[R - 2g^{\mu\nu}\partial_{\mu}\varphi\partial_{\nu}\varphi  - V(\varphi)\right] + S_{\rm matter}(A^2(\varphi) g_{\mu\nu},\chi),
\end{eqnarray}
where $R$ is the scalar curvature with respect to the Einstein frame 
metric $g_{\mu\nu}$, $V(\varphi)$ and $A(\varphi)$ are the potential 
and the coupling function of the scalar field $\varphi$,
and $S_{\rm matter}$ is the action for the matter, 
which is symbolized by $\chi$. 
One should note that an explicit coupling between the matter 
and the scalar field appears through the coupling function $A(\varphi)$ 
only in the Einstein frame due to the conformal transformation 
of the metric involved, while no such direct coupling exists 
in the physical Jordan frame. 
Therefore the weak equivalence principle is satisfied.

In the case of $R^2$ gravity the coupling function $A(\varphi)$ 
and the scalar field potential are given by 
\cite{Yazadjiev:2014cza, Staykov:2014mwa, Yazadjiev:2015zia}
\begin{eqnarray}\label{eq:CouplingFunc}
A(\varphi)=e^{-\frac{1}{\sqrt{3}}\varphi}, \;\; \; 
V(\varphi)= \frac{1}{4a} \left(1-e^{-\frac{2\varphi}{\sqrt{3}}}\right)^2.
\end{eqnarray}
It is straightforward to show that the resulting scalar field is then
massive with a mass $m_\varphi=1/\sqrt{6a}$. 
However, the limiting case $a\rightarrow \infty$ 
corresponds to $m_\varphi=0$, which is equivalent {to a particular class of massless 
Brans-Dicke theories.} 
Since a mass of the scalar field leads to a finite range of the scalar field {of the order of its Compton wavelength}, 
it effectively suppresses the deviations from pure GR. 
Thus the case $a\rightarrow \infty$ 
corresponds to the maximal deviation from GR 
\cite{Yazadjiev:2014cza, Staykov:2014mwa, Yazadjiev:2015zia}. 
On the other hand, in the limit $a\rightarrow 0$ 
the scalar field mass tends to infinity, 
therefore the solutions tend to pure GR solutions.

The free parameter $a$ of the theory is observationally 
constrained to $a \lesssim 10^{11}\, \rm{m}^2$,
which translates in the dimensionless units used in the present paper 
to $a/R_g^2 \lesssim 10^{5}$ \cite{Naf:2010zy}, 
where $R_g = 1.47664\, {\rm km}$ is one half of the solar gravitational radius.
In our calculations we will work with $a$ up to $a\sim 10^5$ 
(in dimensionless units), 
i.e., cover the full observationally allowed range of $a$, 
where the upper limit then leads to the maximal deviations from GR
within this theory.

In the following we will consider static and spherically symmetric solutions. 
Then the Einstein frame metric can be written in the following general form
\begin{eqnarray} 
ds^2 = -e^{2\nu}dt^2 + e^{2\lambda} dr^2 + r^2 (d\theta^2 + \sin^2\theta d\phi^2).
\end{eqnarray}
The dimensionally reduced field equations in the Einstein frame, 
obtained after varying the action (\ref{eq:action}), are given by
\begin{eqnarray} \label{eq:FieldEq}
&&\frac{1}{r^2}\frac{d}{dr}\left[r(1- e^{-2\lambda})\right]= 8\pi G
A^4(\varphi) \bar{\rho} + e^{-2\lambda}\left(\frac{d\varphi}{dr}\right)^2
+ \frac{1}{2} V(\varphi),  \\
&&\frac{2}{r}e^{-2\lambda} \frac{d\nu}{dr} - \frac{1}{r^2}(1-
e^{-2\lambda})= 8\pi G A^4(\varphi) \bar{p} +
e^{-2\lambda}\left(\frac{d\varphi}{dr}\right)^2 - \frac{1}{2}
V(\varphi),\\
&&\frac{d^2\varphi}{dr^2} + \left(\frac{d\nu}{dr} -
\frac{d\lambda}{dr} + \frac{2}{r} \right)\frac{d\varphi}{dr}= 4\pi G
\alpha(\varphi)A^4(\varphi)(\bar{\rho}-3\bar{p})e^{2\lambda} + \frac{1}{4}
\frac{dV(\varphi)}{d\varphi} e^{2\lambda}, \\
&&\frac{d\bar{p}}{dr}= - (\bar{\rho} + \bar{p}) \left(\frac{d\nu}{dr} +
\alpha(\varphi)\frac{d\varphi}{dr} \right),  
\end{eqnarray}
where $\alpha(\varphi)= \frac{d\ln A(\varphi)}{d\varphi}$. 
The quantities $\bar{\rho}$ and $\bar{p}$ are the energy density 
and the pressure in the Jordan frame, and they are connected 
to the respective quantities in the Einstein frame 
via $\rho=A^{4}(\varphi)\bar{\rho}$ and $p=A^{4}(\varphi) \bar{p}$. 

In order to obtain neutron star solutions,
this system of field equations (\ref{eq:FieldEq}) 
has to be supplemented by an equation of state 
for the matter in the Jordan frame of the form $\bar{p}(\bar{\rho})$. 
The boundary conditions are chosen to imply regularity 
at the center of the star and asymptotic flatness at spatial infinity. 
Thus we impose at the center $\bar{\rho}(0)=\bar{\rho}_{c}$, $\lambda(0)=0$ 
and $\frac{d\varphi}{dr}(0) = 0$, where $\bar{\rho}_{c}$ is a free parameter 
denoting the central energy density. 
The condition $\frac{d\varphi}{dr}(0) = 0$ ensures the regularity 
of the corresponding scalar field in both (Einstein and Jordan) frames. 
The Jordan and the Einstein metrics are conformally equivalent, 
and therefore the condition $\lambda(0)=0$ at $r=0$ secures regularity 
of the geometry in both frames. The boundary conditions at infinity, 
which lead to asymptotic flatness in both frames, are 
$\lim_{r\to \infty}\nu(r)=0,$ and $\lim_{r\to \infty}\varphi (r)=0$.

Let us comment on how do we calculate the physical distance 
from the center of the compact object to a fixed point, 
since this will be used later. 
It would be different {from the Einstein frame distance}, 
and it is computed in the following way
\begin{equation} \label{EQ:phys_dist}
r_{\rm phys} = \int_{0}^{r}A(\varphi)e^{\lambda}dr,
\end{equation}
which follows from the fact that the coupling function $A(\varphi)$ 
is actually the conformal factor connecting the metrics of the Jordan frame
and the Einstein frame. 
Another quantity that should be evaluated in the physical Jordan frame 
is the physical radius of the star, which is determined by the requirement 
that the pressure should vanish at the stellar surface $r_S = r(\bar{p}=0)$. 
It is easily seen that the physical radius is given by $R_S = A(\varphi_S)r_S$,
where $\varphi_S$ is the value of the scalar field at the stellar surface. 
The neutron star mass on the other hand is the same in both frames 
\cite{Yazadjiev:2014cza}. Likewise, the frequencies and damping times 
of the QNMs are the same in the Jordan frame and in the Einstein frame.

\subsection{Axial perturbations and exterior complex scaling}

Following the standard procedure 
\cite{Kokkotas:1999bd, Nollert:1999ji, Ferrari:2007dd}
the axial perturbations of the metric can be written as
\begin{eqnarray}
H^{axial}_{\mu\nu} = \left(
\begin{array}{cccc}
0 & 0 &  h_{0}(t,r) S^{lm}_{\theta}(\theta,\phi) & h_{0}(t,r)S^{lm}_{\phi}(\theta,\phi) \\
0 & 0 &  h_{1}(t,r)S^{lm}_{\theta}(\theta,\phi)& h_{1}(t,r)S^{lm}_{\phi}(\theta,\phi) \\
h_{0}(t,r)S^{lm}_{\theta}(\theta,\phi) & h_1(t,r)S^{lm}_{\theta}(\theta,\phi) & 0 & 0 \\
h_{0}(t,r)(r)S^{lm}_{\phi}(\theta,\phi) & h_1(t,r)S^{lm}_{\phi}(\theta,\phi)  & 0 & 0  \\
\end{array}
\right),
\end{eqnarray}
where $(S^{lm}_{\theta}(\theta,\phi),S^{lm}_{\phi}(\theta,\phi))=(-\partial_{\phi}Y_{lm}(\theta,\phi)/\sin\theta,\,\sin\theta \partial_{\theta}Y_{lm}(\theta,\phi))$  with $Y_{lm}(\theta,\phi)$ being the spherical harmonics. Since the scalar field $\varphi$, the pressure $p$ and the energy density $\rho$ transform as true scalars under reflections of the angular coordinates their axial perturbations vanish.      

The relevant perturbed field equations are then given by
\begin{eqnarray} 
&&-e^{-2\nu}\partial^2_{t} h_{1} + e^{-2\nu}\left(\partial_{r} - \frac{2}{r}\right) \partial_{t}h_{0} -\frac{(l-1)(l+2)}{r^2}h_{1}=0 ,\\
&&\partial_{t}h_{0}  - e^{\nu -\lambda}\partial_{r}\left(e^{\nu-\lambda}h_{1}\right)=0.
\end{eqnarray}
Extracting $\partial_{t}h_{0}$ from the second equation 
and substituting it into the first, we find the equation for $h_{1}$. 
But instead of using $h_1$ it is convenient to use the function
\begin{eqnarray}
X= h_1 \frac{e^{\nu-\lambda}}{r}.
\end{eqnarray}  
In terms of the function $X$, the axial perturbations in $f(R)$ gravity are described by the following time-dependent equation 
\begin{eqnarray} \label{EQ:time_pert}
\frac{\partial^2 X}{\partial t^2} &-& e^{\nu-\lambda}\frac{\partial}{\partial r}\left[e^{\nu-\lambda} \frac{\partial X}{\partial r}\right]  + \notag \\
&+&   e^{2\nu}\left[\frac{l(l+1)}{r^2} - \frac{3}{r^2}(1-e^{-2\lambda}) 
+ 4\pi A^4(\varphi)(\bar \rho- \bar p) + \frac{1}{2}V(\varphi) \right] X=0 ,
\end{eqnarray}
where we have also used that  
\begin{eqnarray}
-r e^{-\nu-\lambda} \frac{d}{dr}\left(\frac{e^{\nu-\lambda}}{r^2}\right) = \frac{2}{r^2} - \frac{3}{r^2}\left(1-e^{-2\lambda}\right) +
4\pi G A^4(\varphi)(\bar{\rho} - \bar{p}) + \frac{1}{2}V(\varphi). 
\end{eqnarray}

For the perturbation function $X(r,t)$ 
we can assume the usual time-dependence $X(r,t) = X(r) e^{i\omega t}$,
which leads to a time-independent equation
\begin{eqnarray} \label{EQ:time_ind_pert}
&&e^{\nu-\lambda}\frac{d}{dr}\left[e^{\nu-\lambda} \frac{dX}{dr}\right]  + \notag \\
&&\;\;\;\;\;\;\;\;+ \left\{\omega^2 - e^{2\nu}\left[\frac{l(l+1)}{r^2} - \frac{3}{r^2}(1-e^{-2\lambda}) 
+ 4\pi A^4(\varphi) (\bar \rho - \bar p) + \frac{1}{2}V(\varphi) \right] \right\}X=0.
\end{eqnarray}
The QNM frequency $\omega$ is complex,
$\omega= \omega_R+i \omega_I$,
where the real part $\omega_R$ corresponds to the frequency
of the oscillations, 
while the imaginary part $\omega_I$ is the inverse 
of the damping time $\tau$ of the modes.
  
The perturbation has to be regular at the center of the star, 
which implies $X (r\rightarrow 0) \sim r^{l+1}$. 
At infinity, $r\to\infty$, the general form of the solution 
is a linear combination of an outgoing and an ingoing wave
\begin{eqnarray} %
X \sim A_{in} e^{i\omega (t+R)} + A_{out} e^{i\omega (t-R)},
\end{eqnarray}
where $dR=e^{\lambda-\nu} dr$ defines the tortoise coordinate. 
Since we are interested in neutron star QNMs,  
$X$ should have the form of a purely outgoing wave at infinity 
without any ingoing wave contribution. 
However, trying to impose this condition numerically
leads to severe numerical problems. 
For example in the case of stable modes with $\omega_I>0$ 
the outgoing perturbation diverges for large distance,
while the ingoing contribution goes exponentially to zero.
Thus, any small numerical contamination with an ingoing wave contribution 
leads to errors in the determination of the resonant frequencies.  

This problem of course exists only when we solve the time-independent 
perturbation equation \eqref{EQ:time_ind_pert}. 
If we perform a direct time evolution of the time-dependent 
perturbation equation \eqref{EQ:time_pert}, 
the outgoing wave condition does not have to be imposed explicitly. 
Instead, we should just locate the rightmost boundary  
of the integration domain far enough from the center of the star,
so that any ingoing contamination of the signal would have to travel 
too long to the point of extraction of the gravitational wave signal
and, thus, would not influence the final results.

In order to control the error coming from the numerical contamination 
with an ingoing wave when calculating the resonant frequencies 
using the time-independent perturbation equation \eqref{EQ:time_ind_pert}, 
we employ the method based on exterior complex scaling (see e.g.,
\cite{PhysRevA.44.3060,BlazquezSalcedo:2012pd,Blazquez-Salcedo:2013jka,Blazquez-Salcedo:2015ets}).
Instead of integrating directly Eq.~(\ref{EQ:time_ind_pert})
for the perturbation $X$, 
we consider the phase function of the perturbation. 
Thus we transform Eq.~(\ref{EQ:time_ind_pert}) 
to the form of a Riccati equation \cite{Chandrasekhar:1975zza}
by defining the phase function of the perturbation 
via $\frac{dX}{dr} = g X$. Then we obtain
\begin{eqnarray} \label{EQ:phase}
\frac{dg}{dr}=-g^2 - e^{2\lambda-2\nu}\omega^2 
+g e^{2\lambda} \left[  4\pi r A^4(\varphi)({\bar \rho}-{\bar p}) +\frac{r}{2}V(\varphi) - \frac{1}{r} \left(1-e^{-2\lambda}\right) \right] \nonumber \\+ e^{2\lambda} \left[  4\pi A^4(\varphi) (\bar\rho-\bar p) + \frac{1}{2}V(\varphi) + \frac{l(l+1)}{r^2 }- \frac{3}{r^2}\left(1-e^{-2\lambda}\right) \right] .
\end{eqnarray}
Regular perturbations at the center of the neutron star 
will possess a phase function behaving like $g \sim \frac{l+1}{r}$ 
close to $r=0$. 
At infinity, however, a general solution of the equation will look like
\begin{eqnarray} \label{phase_inf}
g \sim i\omega \frac{A_{in} e^{i\omega r} - A_{out} e^{-i\omega r}}{A_{in} e^{i\omega r} + A_{out} e^{-i\omega r}},
\end{eqnarray}
which tends to $g|_{\infty} = -i\omega$ asymptotically, 
for both purely outgoing waves and mixed (ingoing plus outgoing) waves.

In order to guarantee that the solutions correspond to purely outgoing waves, 
we make an analytical continuation of the equation into the complex plane. 
This can be done by defining a complex coordinate $r$ 
\cite{Andersson:1997xt}
\begin{eqnarray} %\label{phase_inf}
r = r_j + y e^{-i\zeta},
\end{eqnarray}
where the real numbers $r_j$ and $\zeta$ are auxiliary constants, 
and $y \in [0,\infty)$. The asymptotic behaviour of the phase function $g$
then becomes for $y \to \infty$
\begin{eqnarray} \label{phase_2_inf}
g \sim i\omega \frac{A_{in} e^{i\xi y} - A_{out} e^{-i\xi y}}{A_{in} e^{i\xi y} + A_{out} e^{-i\xi y}},
\end{eqnarray}
with $\xi = \xi_R + i \xi_I$ 
and $\xi_R = \omega_R\cos\zeta + \omega_I\sin\zeta$, 
$\xi_I = \omega_I\cos\zeta -\omega_R\sin\zeta$. 
If we now choose $\zeta$ such that
$\xi_I = \omega_I\cos\zeta -\omega_R\sin\zeta < 0$, 
the condition $g|_{\infty} = -i\omega$ enforces $A_{in}=0$, 
and the solution will describe purely outgoing waves.

With this setting, the resulting numerical procedure is the following. 
First we generate a static background solution, where we have 
to fix the parameters of the theory (i.e., the constant $a$), 
the equation of state, and the central pressure of the configuration. 
In the next step, we calculate the coefficients of Eq.~(\ref{EQ:phase}) 
using the background configuration. 
Then we generate an \textit{interior} solution of this equation 
extending from the center of the star at $r=0$, up to a point at $r=r_j$,
where the point $r_j$ is chosen to be outside the star ($r_j > R_s$),
and we demand regularity of the phase function by imposing 
$\lim_{r\to 0} r \cdot g = l+1$. 

Next, a second solution of Eq.~(\ref{EQ:phase}) is generated, 
but this time after performing the transformation (\ref{phase_inf}) 
with $\omega_I\cos\zeta -\omega_R\sin\zeta < 0$. 
This solution extends from $r=r_j$ to infinity, 
and we demand the \textit{exterior} phase to satisfy 
$g|_{\infty} = -i\omega$, which imposes the outgoing wave behaviour.
QNMs are obtained when the interior phase solution 
and the exterior phase solution are continuous at the point $r=r_j$. 
In our case, this only happens for a discrete set of values of $\omega$. 
We implement a search algorithm that minimizes the difference 
between both solutions at $r=r_j$ using the gradient descent. 
Once we obtain a QNM, we test the numerical stability 
by changing the auxiliary parameters $r_j$ and $\zeta$. 
Typically the QNMs can be obtained with a relative accuracy of $10^{-3}$ 
or better.

\section{Numerical results}

We used two independent numerical techniques for solving the problem 
in order to verify the results and gain deeper insight. 
First we performed a time evolution of Eq.~(\ref{EQ:time_pert}) 
perturbing the star with some initial pulse. 
This is the most general way of constructing a solution 
since we just impose some initial perturbation, 
and then let the system evolve. The resulting signal will contain
an admixture of all the excited modes. 
The drawback of this method is that it normally yields only 
the first axial QNM because of the short damping times 
of the higher axial modes. 
In addition, only a few oscillations can be observed 
before the signal dies out due to the fast damping, 
which leads to large errors in the extracted frequencies and damping times.

Solving the time-independent perturbation equation (\ref{EQ:time_ind_pert}) 
via the complex scaling method is of course more involved,
but it is a much more robust procedure,
that can be used to explore the parameter space with high accuracy. 
Therefore, we used this method to obtain the QNMs of a large variety of EOS 
and extract a set of universal relations. 
The background solutions were obtained 
using the ODE solver Colsys \cite{Ascher:1979iha}.
Typically, the mesh sizes of the solutions include $10^4$-$10^5$ points, 
with a relative accuracy of $10^{-10}$ for both functions and derivatives. 
To implement the realistic equations of state, 
we used a monotonic cubic Hermite interpolation of the data points 
for the tabulated EOS, or alternatively, 
a piece-wise polytropic interpolation as described in \cite{Read:2008iy}.
The results obtained with the two methods 
proved to be in very good agreement,
pointing to the correctness of the calculations.

\subsection{Time evolution of the axial perturbation equation}

We start our discussion of the numerical results 
by examining the axial oscillations of neutron stars 
in GR and $f(R)$ gravity via time evolution 
of the time-dependent equation (\ref{EQ:time_pert}). 
We here concentrate mainly on the characteristics 
of the problem that can be extracted solely from the time evolution 
including a comparison with the results from the complex scaling method, 
and leave the detailed discussion of the behavior of the 
oscillation frequencies, damping times and universal relations 
for the next sections. 

{We evolved Eq.~(\ref{EQ:time_pert})} for a given period of time 
imposing a Gaussian pulse as initial data\footnote{We
used a Gaussian pulse for convenience.
The results for the axial mode frequencies should be independent of 
the form of the initial pulse.}.
The frequency was calculated from the obtained waveform 
via Fourier transformation of the signal. 
The damping times were obtained by fitting the peaks of the signal  
with an exponential function. 
The coefficient in this fitting exponent is 
the imaginary part of the frequency, which is
related to the damping time via $\tau = 1/\omega_{I}$.

We studied neutron star models with a realistic EOS, 
namely the APR4 EOS \cite{Akmal:1998cf}.
The presented waveforms were extracted at about 200 km physical distance 
from the star as defined according to Eq.~\eqref{EQ:phys_dist}. 
This distance was chosen in such a way that it is on one hand far enough 
from the star, and on the other hand the outer boundary of the numerical 
integration was located at much larger radial distances,
so that any contamination from ingoing waves 
(due to the finiteness of the computational domain) was avoided. 

The results we present in this subsection are for 
very massive neutron stars only. 
The reason is that the damping time of the considered modes 
increases with the increase of the compactness 
which leads to a larger number of oscillations to be observed 
before the signal dies out. 
Since the axial modes have very small damping times in general, 
the number of oscillations is not that large even for the most massive models. 
For the considered EOS, one can determine the oscillation frequencies 
and damping times with good accuracy only for masses of the order 
of two solar masses. Since one of the main motivations 
for performing time evolution calculations is to verify the results 
from solving the time-independent problem via the complex scaling method, 
we have chosen to work with masses of $2M_\odot$ and above
in order to have high enough accuracy.

\begin{figure}[]
	\centering
	\includegraphics[width=0.45\textwidth]{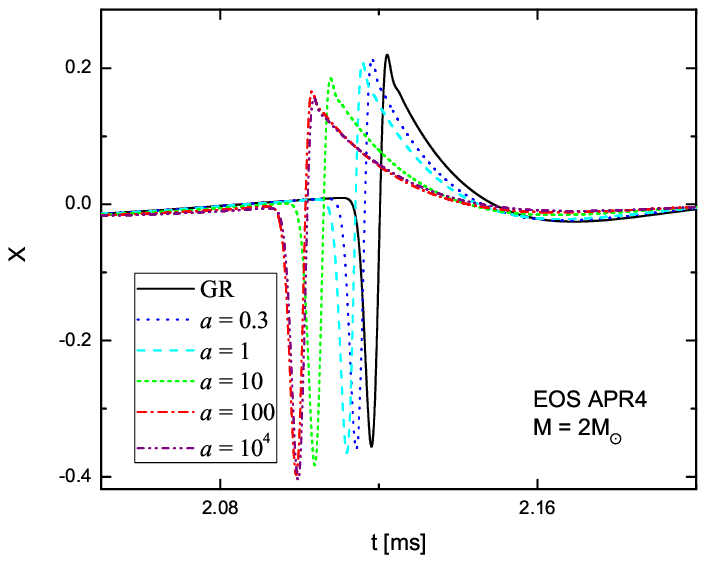}
	\includegraphics[width=0.45\textwidth]{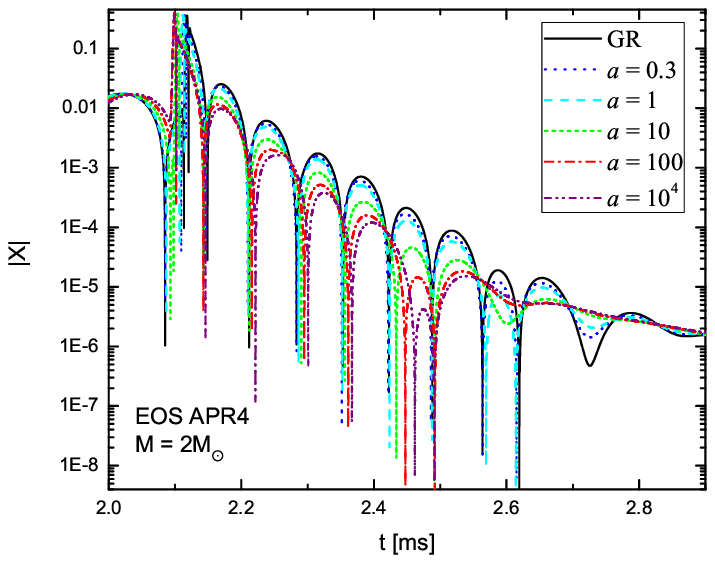}
	\caption{The observed signal at approximately 200 km physical distance 
from the star obtained via time evolution of the time-dependent 
perturbation equation. The studied neutron star models are obtained 
with EOS APR4, mass $M = 2 M_{\odot}$ and for different values
of the parameter $a$ between GR (black) and $a=10^4$ (purple).
(\textit{left})
The signal is presented as a function of time in milliseconds. 
(\textit{right}) 
The absolute value of the 
same signal is shown on a logarithmic scale as a function of time.}
	\label{Fig:APR4_2M}
\end{figure}

\begin{figure}[]
	\centering
	\includegraphics[width=0.45\textwidth]{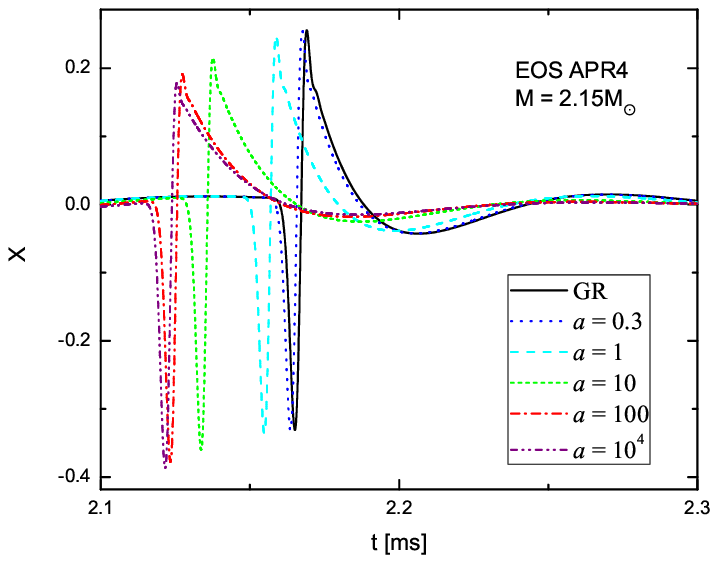}
	\includegraphics[width=0.45\textwidth]{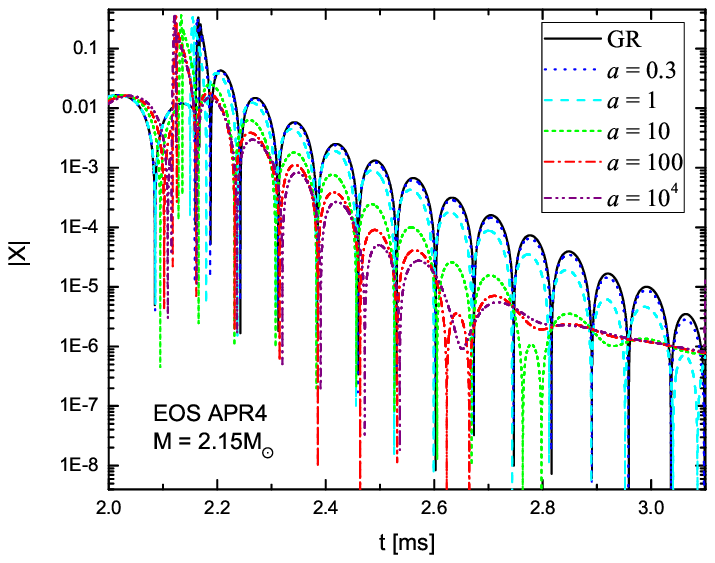}
	\caption{The observed signal at approximately 200 km physical distance 
from the star obtained via time evolution of the time-dependent 
perturbation equation. The studied neutron star models are obtained
with EOS APR4, mass $M = 2.15 M_{\odot}$ and for different values
of the parameter $a$ between GR (black) and $a=10^4$ (purple).
(\textit{left})
The signal is presented as a function of time in milliseconds.
(\textit{right}) 
The absolute value of the
same signal is shown on a logarithmic scale as a function of time.}
	\label{Fig:APR4_2.15M}
\end{figure}

The waveforms for neutron star models with masses $M=2.0M_\odot$ 
and $M=2.15M_\odot$ and for the fundamental $l=2$ curvature mode 
are presented in Figs.~\ref{Fig:APR4_2M} and \ref{Fig:APR4_2.15M}. 
In the left panels of both figures the signal is presented 
as a function of time, and in the right panels the absolute value 
of the signal is shown on a logarthmic scale as a function of time. 
As commented, the signal is extracted at {a physical distance} from the star 
which is roughly equal to 200 km. 

The figures show the dependence of the arrival time of the signal 
on the free parameter $a$. As expected, the maximal deviation 
from GR, and hence the shortest arrival time, 
occurs for the largest value of $a$ considered,
and the arrival time gets closer to the GR value when $a$ decreases. 
This effect arises from two sides.
First, we extract the signal at a fixed physical distance, 
that differs from the coordinate one according to Eq.~\eqref{EQ:phys_dist}. 
Second, the effective potential of the 
perturbation equation \eqref{EQ:time_ind_pert} is influenced 
by the presence of a scalar field (or $R^2$ modification of GR). 
A similar effect of a delay in the arrival time of the signal 
is observed also in \cite{Sotani:2005qx} 
for the case of scalar-tensor theories with a massless scalar field.  

In Table \ref{Tbl:TimeEvol} we present the oscillation frequencies 
and damping times extracted from the time evolution results presented 
in Figs.~\ref{Fig:APR4_2M} and \ref{Fig:APR4_2.15M} together with
the results obtained by solving the boundary value problem.
As one can see there is a good agreement between the results 
for the oscillation frequencies and damping times obtained with both methods. 
The maximal deviation between both methods is about 3\% 
for all of the considered models\footnote{This deviation might increase 
with the decrease of the stellar compactness due to the low accuracy 
of the results coming from the decrease of the damping time. 
That is why we have chosen to work with higher mass models 
as commented above.}.  

The data in the table show that both the frequencies of the modes 
and the damping times decrease with increasing $a$. 
This leads to a serious decrease of the number of oscillations 
observed in the signal in Figs.~\ref{Fig:APR4_2M} and \ref{Fig:APR4_2.15M} 
for large $a$ due to the fast damping. 
We leave further comments on the behavior of these quantities 
for the following subsections.

\begin{center}
\begin{table}[h]
\begin{tabular}{lcccc}
Model & Time evolution & BVP & Time evolution & BVP \\
\quad & $\omega_R$ [kHz] & $\omega_R$ [kHz] & $\tau$ [$\mu$s] & $\tau$ [$\mu$s] \\
\hline
GR          & 7.222  & 7.300  & 61.999  & 62.618 \\
$a = 0.3$   & 7.259  & 7.277  & 61.194  & 61.637 \\
$a = 1$     & 7.175  & 7.232  & 59.265  & 59.728 \\
$a = 10$    & 6.909  & 7.049  & 54.612  & 54.291 \\
$a = 10^2$  & 6.820  & 6.891  & 51.902  & 50.770 \\
$a = 10^4$  & 6.578  & 6.803  & 48.266  & 49.034 \\
\end{tabular}
\caption{The oscillation frequencies and the damping times of neutron stars 
with EOS APR4, mass $M = 2 M_{\odot}$ and for different values
of the parameter $a$ between GR and $a=10^4$.
For comparison results obtained with the time evolution 
and the time-independent boundary value problem (BVP) are presented.}
\label{Tbl:TimeEvol}
\end{table}
\end{center}

\subsection{Boundary value problem}

\subsubsection{Axial quasi-normal mode spectrum for different EOS and values of $a$}

In this subsection we present the dependence of the spectrum of QNMs 
on the parameter $a$ obtained by solving the time-independent 
perturbation equation \eqref{EQ:time_ind_pert} via the complex scaling method. 
We consider a set of ten realistic equations of state 
describing different matter contents. 
To label the EOS we follow the nomenclature used previously in 
\cite{Blazquez-Salcedo:2015ets,BlazquezSalcedo:2012pd,Blazquez-Salcedo:2013jka,Motahar:2017blm}. 
Some of the EOS include purely nuclear matter (SLy, APR4), 
others hyperon matter (GNH3, H4, BHZBM, WCS1-2), 
or a mixture of quark and nuclear matter (ALF2-4, WSHPS3). 
We now focus on the fundamental $l=2$ curvature mode.

\begin{figure}[]
	\centering
	\includegraphics[width=0.37\textwidth,angle=-90]{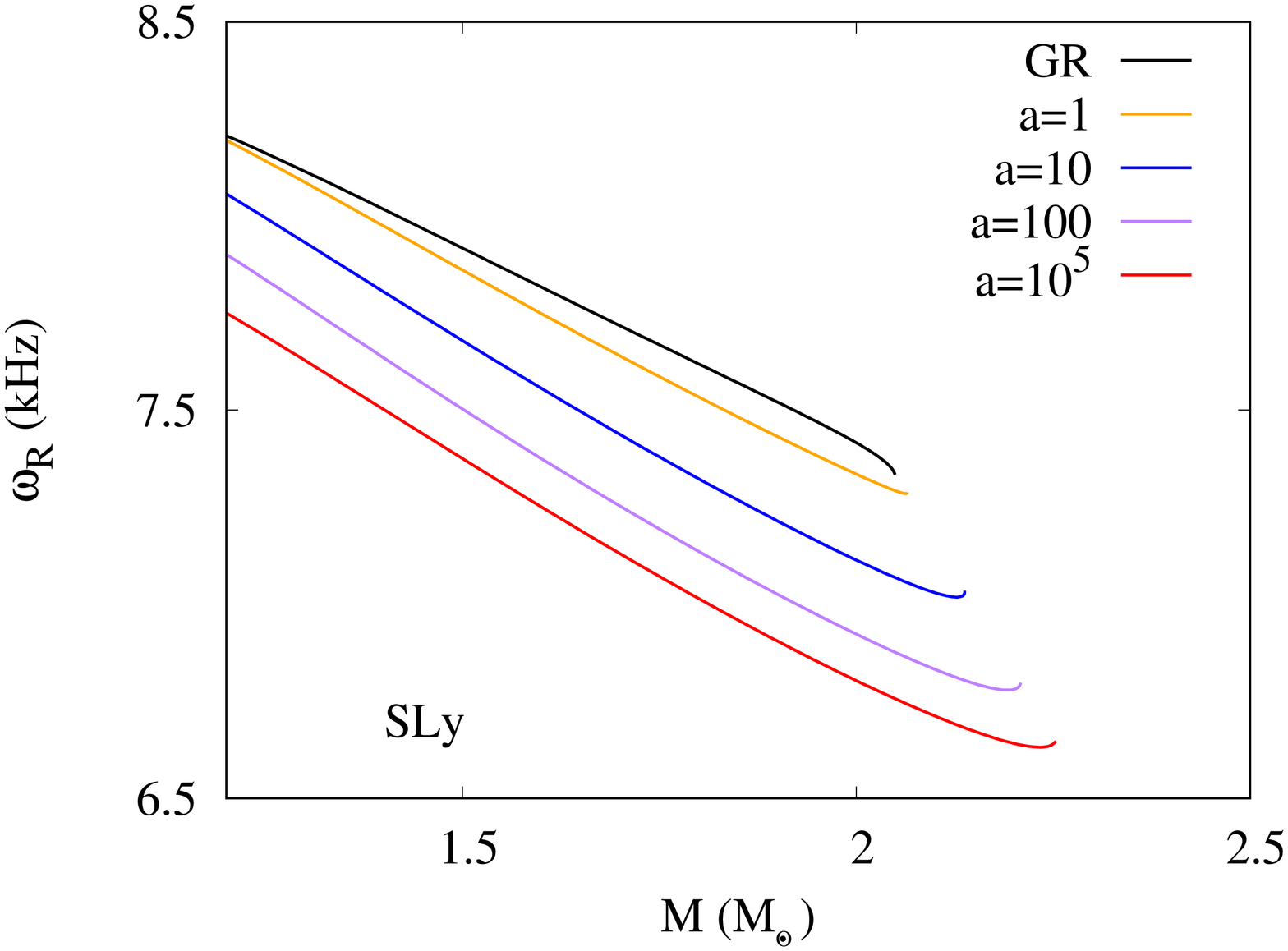}
	\includegraphics[width=0.37\textwidth,angle=-90]{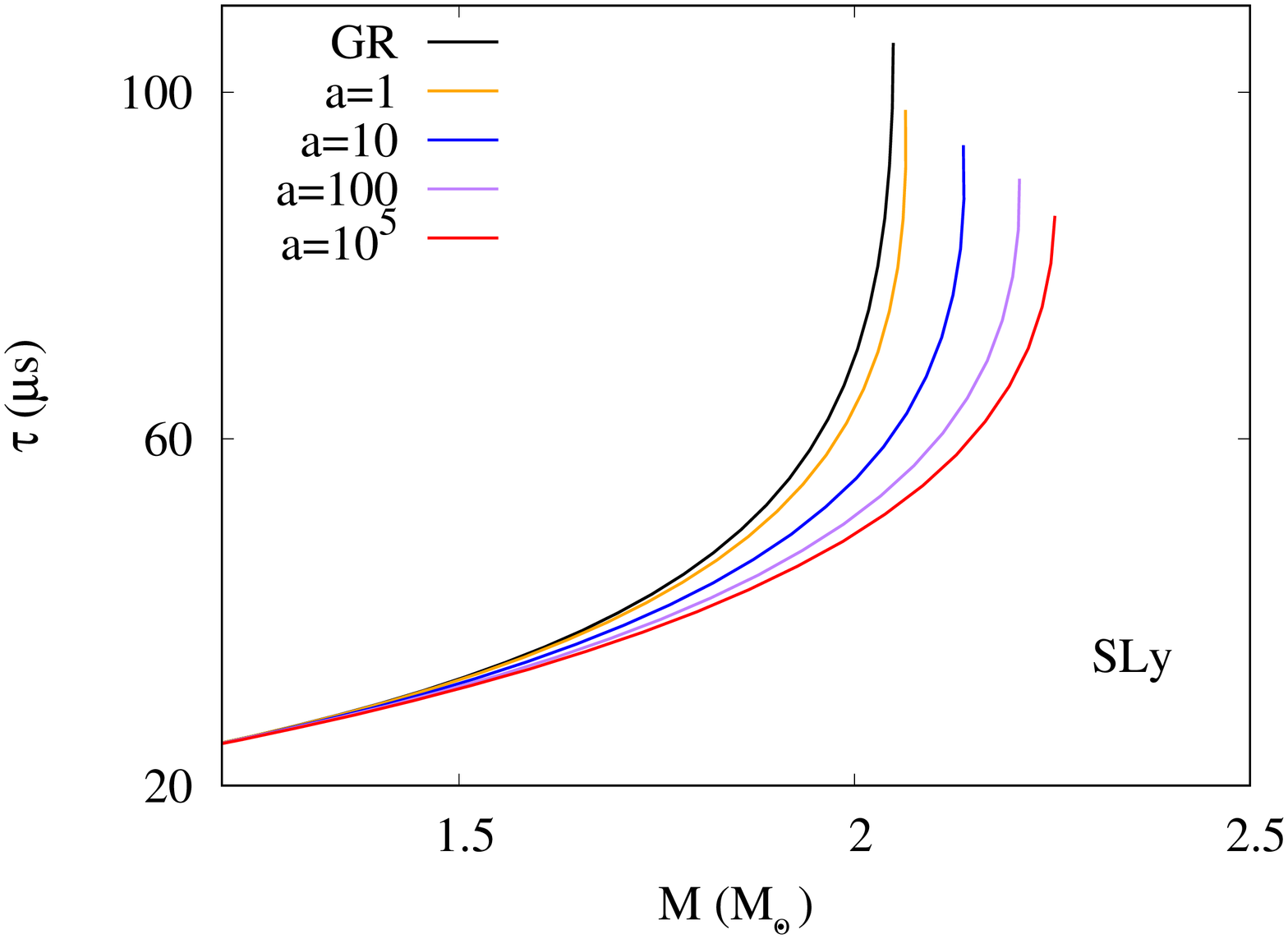}
	\caption{(\textit{left}) Frequency $\omega_R$ (in units of kHz) 
of the fundamental $l=2$ curvature mode versus the mass 
(in units of the solar mass $M_\odot$) of neutron stars 
with the SLy equation of state and different values 
of the parameter $a$ varying between GR (black) and $a=10^5$ (red). 
(\textit{right}) A similar figure showing the damping time $\tau$ (in $\mu s$). 
	}
	\label{l2_Sly}
\end{figure}

\begin{figure}[]
	\centering
	\includegraphics[width=0.9\textwidth,angle=-90]{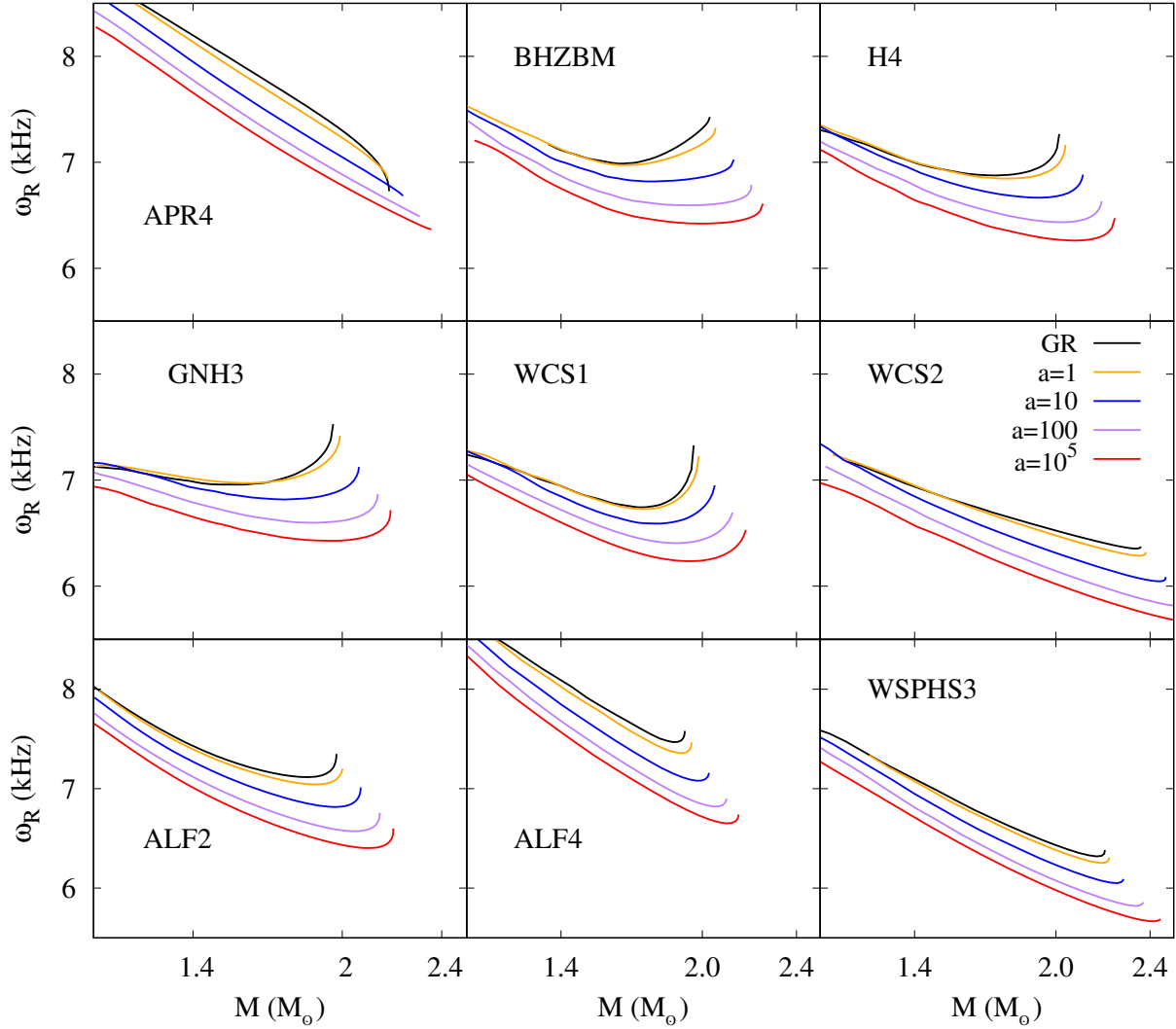}
	\caption{Frequency $\omega_R$ (in units of kHz)
of the fundamental $l=2$ curvature mode versus the mass
(in units of the solar mass $M_\odot$) of neutron stars
for different values
of the parameter $a$ between GR (black) and $a=10^5$ (red).
Each panel corresponds to a different equation of state.}
	\label{l2_omegaR_9EOS}
\end{figure}

\begin{figure}[]
	\centering
	\includegraphics[width=0.9\textwidth,angle=-90]{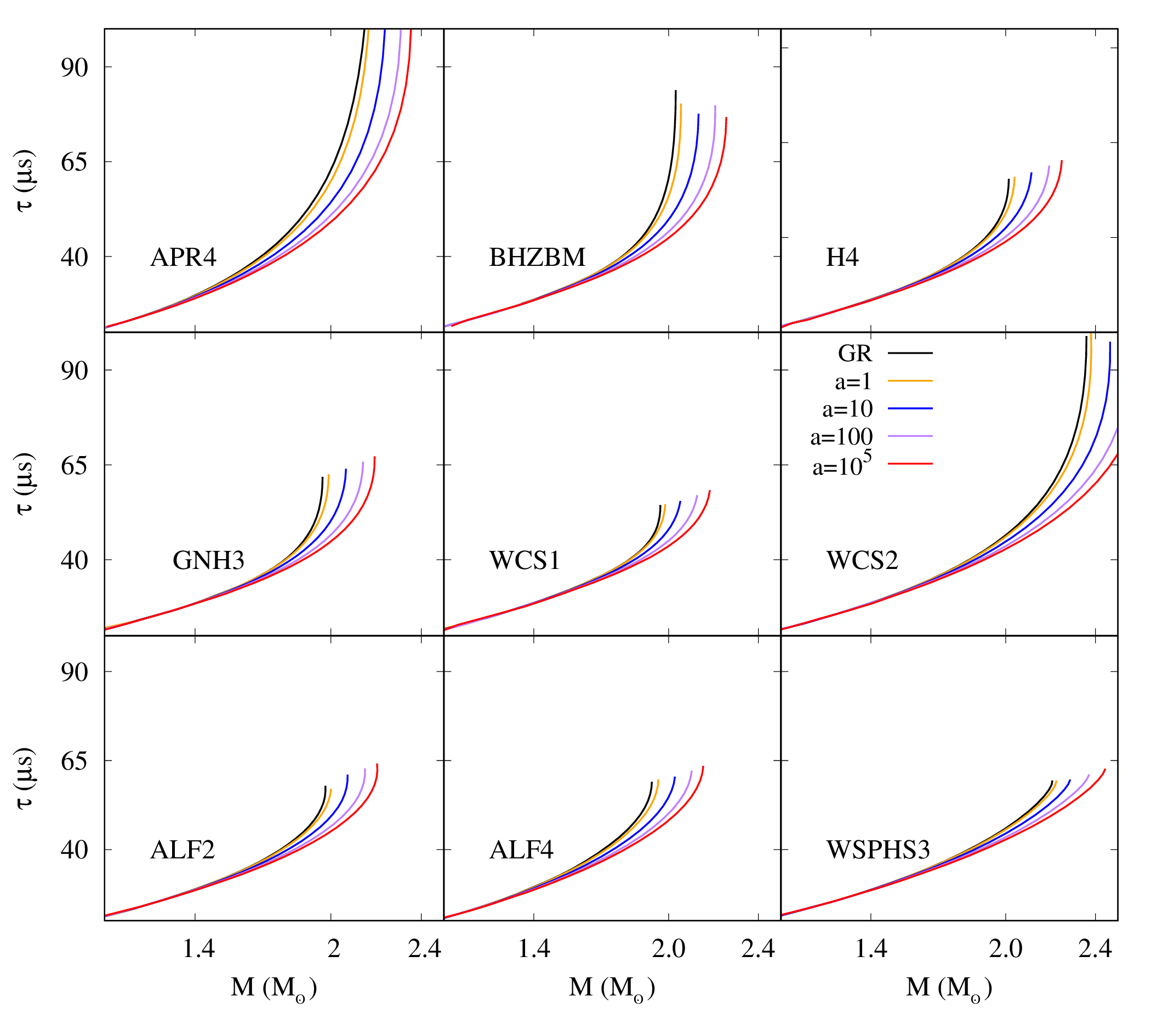}
	\caption{Damping time $\tau$ (in units of $\mu$s)
of the fundamental $l=2$ curvature mode versus the mass
(in units of the solar mass $M_\odot$) of neutron stars
for different values
of the parameter $a$ between GR (black) and $a=10^5$ (red).
Each panel corresponds to a different equation of state.}
	\label{l2_omegaI_9EOS}
\end{figure}

As an example of the typical effect of the $R^2$ term on the QNMs, 
we present in Fig.~\ref{l2_Sly} the fundamental $l=2$ curvature mode 
for the SLy EOS. In Fig.~\ref{l2_Sly} (left panel) 
we show the frequency $\omega_I$ in $\rm kHz$ 
versus the total mass of the neutron star in $M_\odot$. 
In black we show the frequencies for the GR case 
and in red for the $a=10^5$ case, 
and in between for further values of the parameter $a$  
(in orange $a=1$, in blue $a=10$, and in purple $a=100$). 
In general the frequency is maximal in GR
and decreases as the value of $a$ is increased, 
becoming minimal for the {massless Brans-Dicke theory}. 
Note that since the maximum mass increases with the value of $a$, 
the lowest frequencies are reached by {the massless Brans-Dicke neutron stars} 
with masses close to the maximum.
Similarly, in Fig. \ref{l2_Sly} (right panel) 
we show the damping time $\tau$ in $\rm \mu s$ as a function of the total mass.
The damping time is largest in GR, and decreases with increasing value of $a$. 
Interestingly, for low mass configurations the values of the damping time 
become rather independent of $a$. 
The damping time increases monotonically with the mass in all cases. 
These features are present independently of the equation of state. 
In each panel of Fig.~\ref{l2_omegaR_9EOS} we show the frequency $\omega_R$
as a function of the total mass for a different EOS and several values of $a$. 
Similarly in Fig.~\ref{l2_omegaI_9EOS} we show the damping time $\tau$.

\begin{figure}[]
	\centering
	\includegraphics[width=0.37\textwidth,angle=-90]{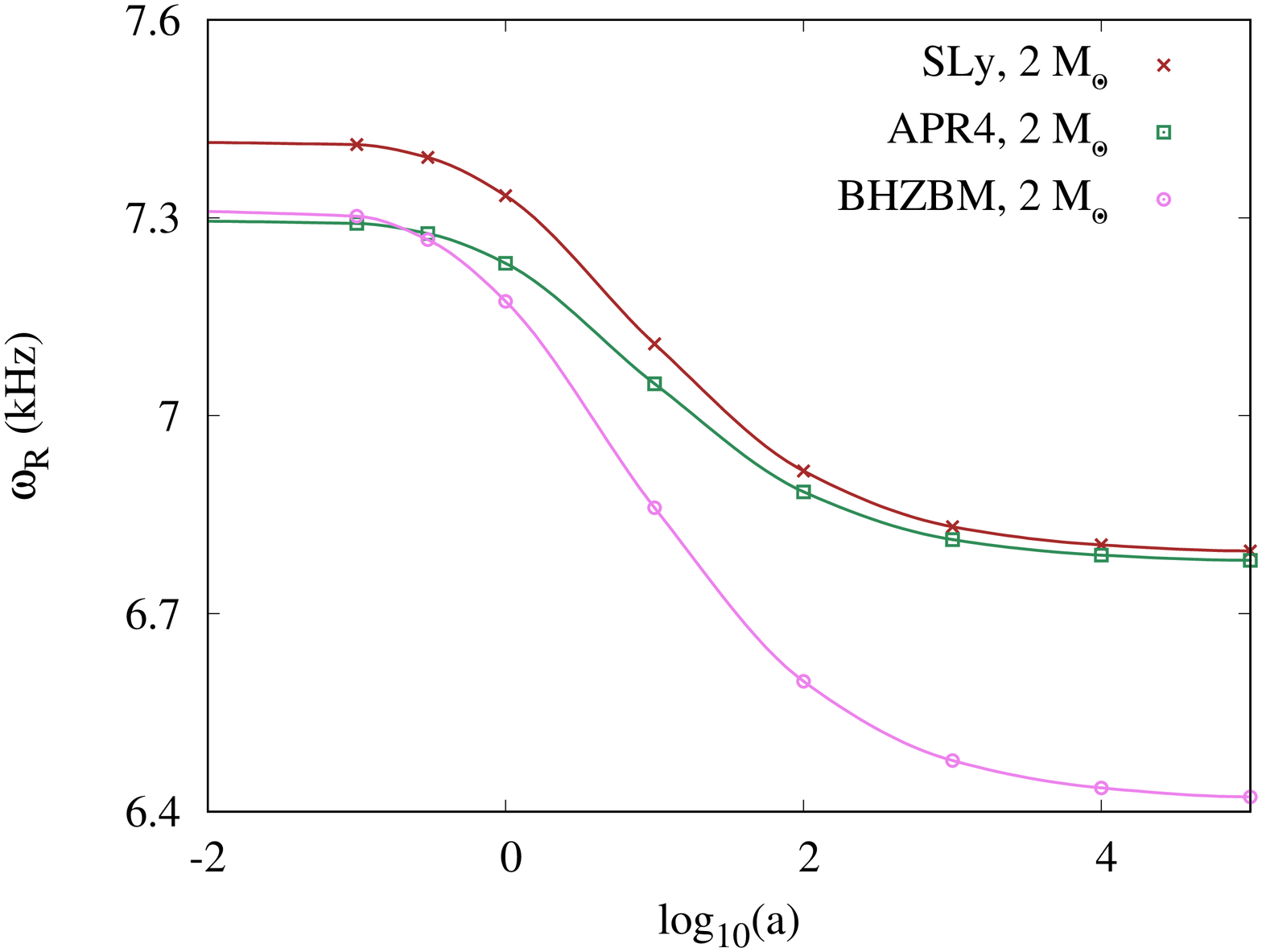}\label{l2_2M_real}
	\includegraphics[width=0.37\textwidth,angle=-90]{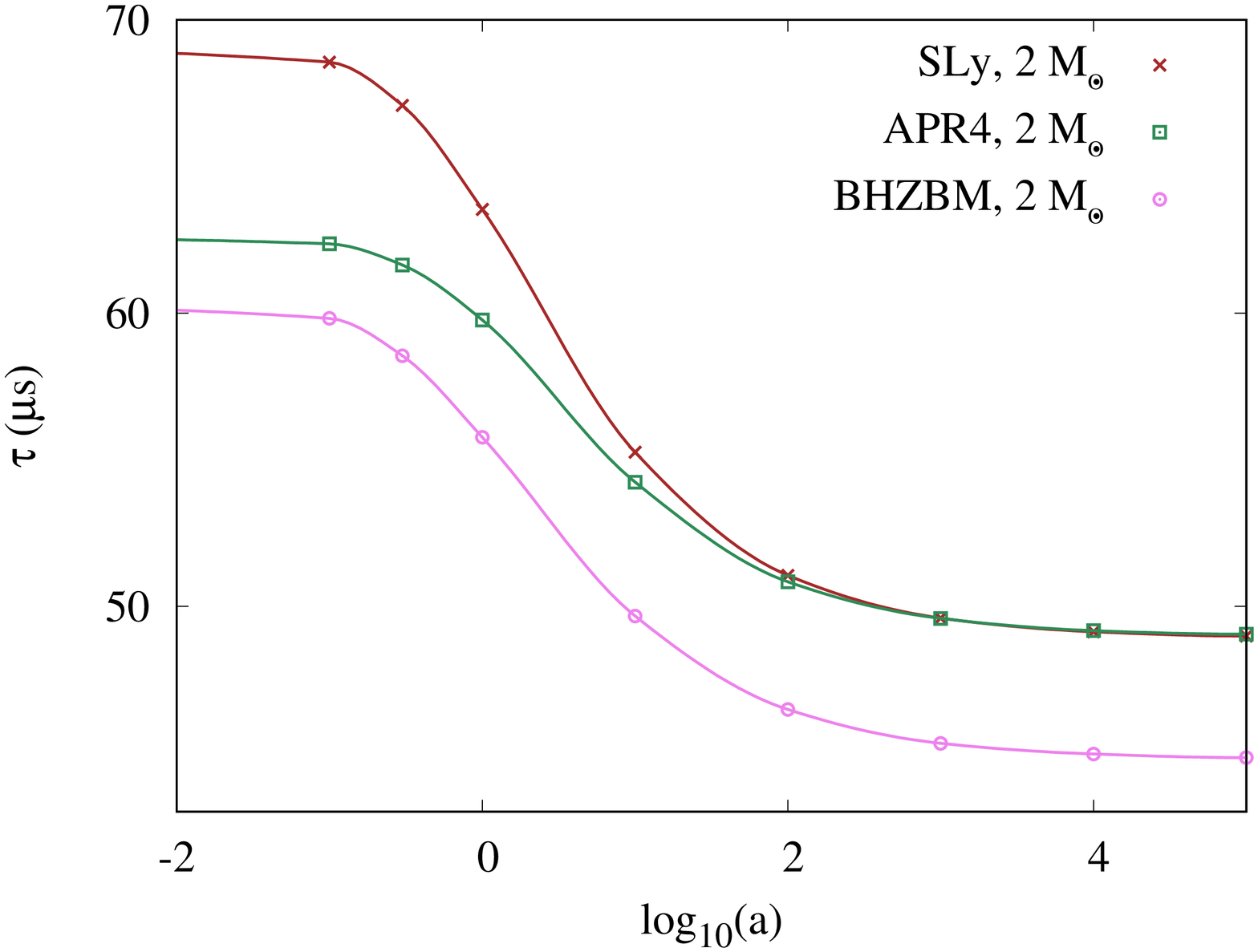}\label{l2_2M_img}
	\caption{(\textit{left}) 
Frequency $\omega_R$ (in units of kHz)
of the fundamental $l=2$ curvature mode versus the parameter $a$
on a logarithmic scale for a neutron star mass of 2 $M_\odot$
and three EOS (Sly brown, APR4 green, BHZBM purple).
(\textit{right}) A similar figure for the damping time $\tau$ ($\rm \mu s$).}
	\label{l2_2M}
\end{figure}

In Fig.~\ref{l2_2M} we show the frequency $\omega_R$ (left panel) 
and the damping time $\tau$ (right panel) as a function of the parameter $a$ 
for several EOS (SLy in brown, APR4 in green and BHZBM in purple) 
and for a fixed value of the total mass, $M=2 M_{\odot}$. 
Both the frequency and the damping time decrease monotonically 
as the value of $a$ increases. 
This behaviour is observed for large values of the mass 
(i.e., larger compactness). 
However, for small values of the mass (around $1 M_{\odot}$), 
this behaviour can change for some equations of state, 
with the frequency slightly increasing for small values of $a$, 
as can be seen in Fig.~\ref{l2_omegaR_9EOS} (e.g., for GNH3 and WCS1). 
The limit $a\to 0$ leads to the corresponding GR value of the mode, 
while the limit $a\to \infty$ gives {the massless Brans-Dicke value}. 

\subsubsection{Universal relations}

\begin{figure}[]
	\centering
	\includegraphics[width=0.49\textwidth,angle=-90]{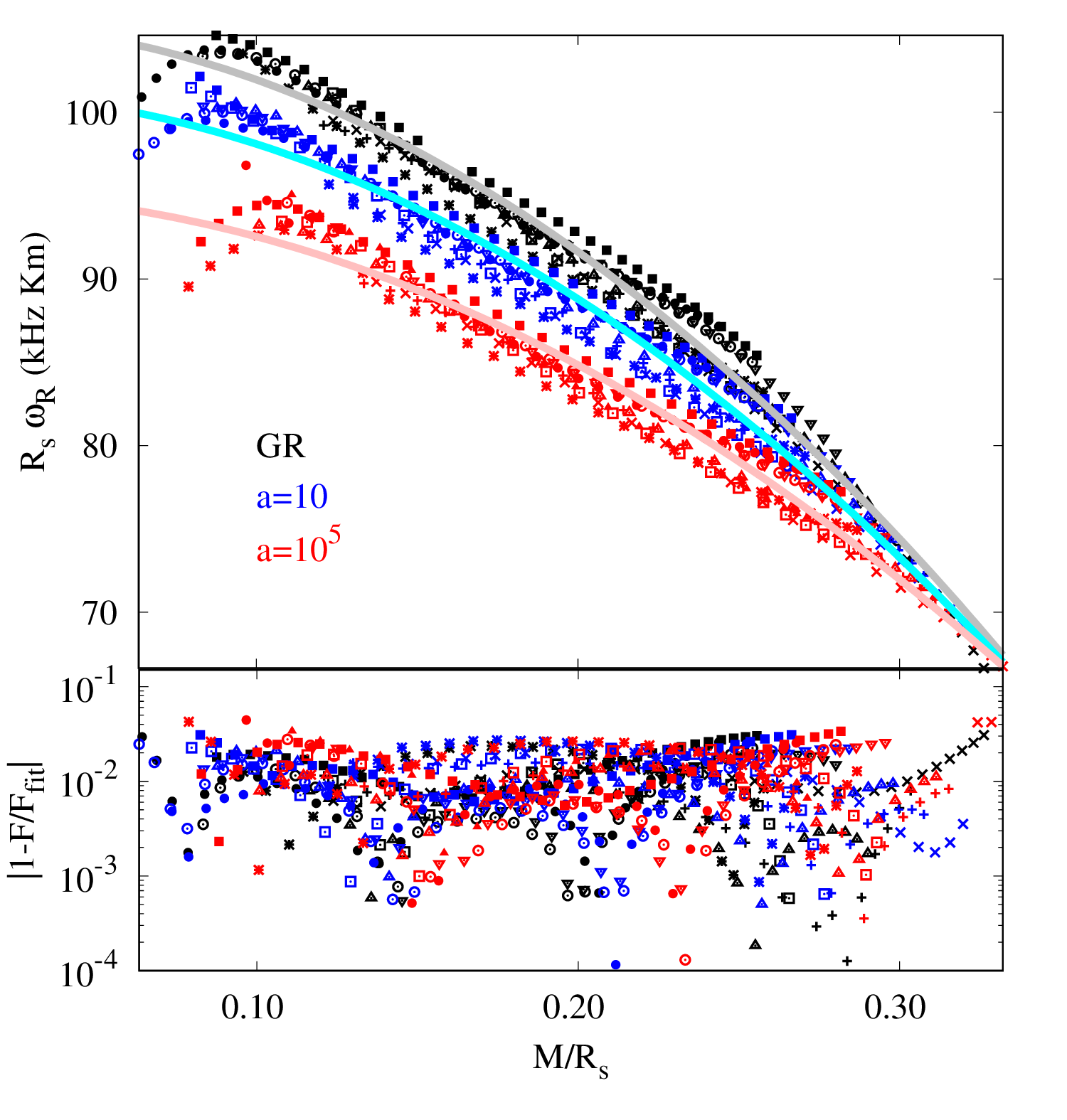}%\label{l2_UR_real}
	\includegraphics[width=0.49\textwidth,angle=-90]{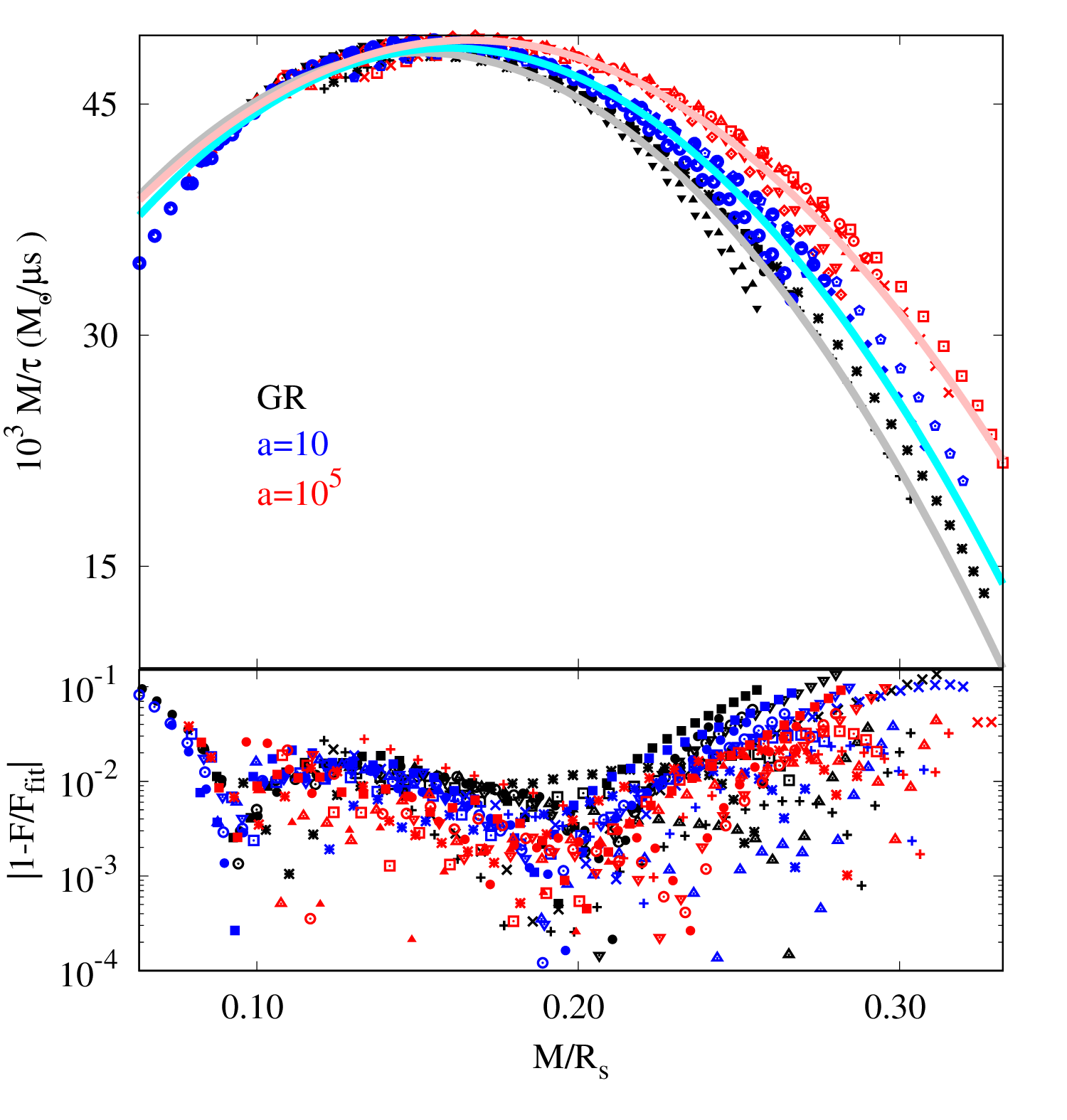}%\label{l2_UR_img}
	\caption{(\textit{left}) Frequency $\omega_R$ scaled with the radius 
$R_s$ of the star ($\rm kHz \cdot km$) versus compactness $M/R_s$. 
(\textit{right}) Inverse of the damping time $\tau$ scaled 
with respect to the mass of the star ($M_{\odot}/\mu \rm s$) 
versus compactness $M/R_s$
in black for GR, in blue for $a=10$, and in red for $a=10^5$.
The solid lines correspond to the fits (\ref{fit_freq}) and (\ref{fit_tau}) 
(in grey for GR, in cyan for $a=10$ and in pink for $a=10^5$).}
	\label{l2_UR}
\end{figure}

Let us now address (approximate) universal relations 
for the rescaled frequencies and the rescaled damping times  
\cite{Kokkotas:1999mn,Blazquez-Salcedo:2015ets,Benhar:2004xg,Andersson:1996pn,Andersson:1997rn}.
In Fig.~\ref{l2_UR} (left panel) we exhibit the frequency $\omega_R$ 
rescaled with the radius $R_s$ of the star versus the compactness 
$M/R_s$ of the star. The results are shown for GR, and for
$f(R)$ theory with $a=10$ and $a=10^5$. 
The three sets of results are fitted with the same phenomenological relation, 
where different coefficients are obtained, namely 
\begin{eqnarray} 
\omega_R[\rm kHz]\cdot R_s [\rm km] = 
\begin{cases}
(105.39\pm 0.14) - (344.2\pm 2.8) \left( \frac{M}{R_s} \right) ^2 &\textnormal{GR}\\
(101.19\pm 0.14) - (310.0\pm 3.0) \left( \frac{M}{R_s} \right) ^2 &a=10\\
(53.20\pm 0.42) - (203.7\pm 7.8) \left( \frac{M}{R_s} \right) ^2 &a=10^5
\end{cases}
\label{fit_freq}
\end{eqnarray}
In all cases the deviation from universality (i.e., the best fit)
is less than 5\%, as seen in the bottom panel of the figure. 
As expected, the normalized frequency is highest for GR,
decreases with increasing $a$, 
and {is smallest for $a=10^5$}.

In Fig.~\ref{l2_UR} (right panel) we exhibit the inverse 
of the damping time $\tau$ rescaled with the mass of the star versus 
the compactness of the models. 
Again, the phenomenological relation has the same 
form as in GR but with different coefficients 
\begin{eqnarray}
10^3 \frac{M [M_{\odot}]}{\tau [\rm \mu s]} = 
\begin{cases}
(20.80\pm 0.56) + (365.6\pm 6.0) \frac{M}{R_s} - (1213\pm 15) \left( \frac{M}{R_s} \right) ^2 &\textnormal{GR}\\
(18.78\pm 0.45) + (374.0\pm 5.0) \frac{M}{R_s} - (1170\pm 13) \left( \frac{M}{R_s} \right) ^2 &a=10\\
(21.99\pm 0.53) + (327.4\pm 5.5) \frac{M}{R_s} - (986\pm 13) \left( \frac{M}{R_s} \right) ^2 &a=10^5
\end{cases}
\label{fit_tau}
\end{eqnarray}
The deviations from universality are higher in this case 
with models for maximal compactness showing deviations 
of up to about 10\%. 
The lowest normalized inverse damping times are present for GR,
and {they increase with the increase of the parameter $a$.}
Both phenomenological relations show significant maximal deviations 
from GR in different parts of the compactness scale,
which may prove quite useful for deriving constraints on the parameter 
of the theory, once the observations will reach the necessary accuracy.  

\begin{figure}[]
	\centering
	\includegraphics[width=0.49\textwidth,angle=-90]{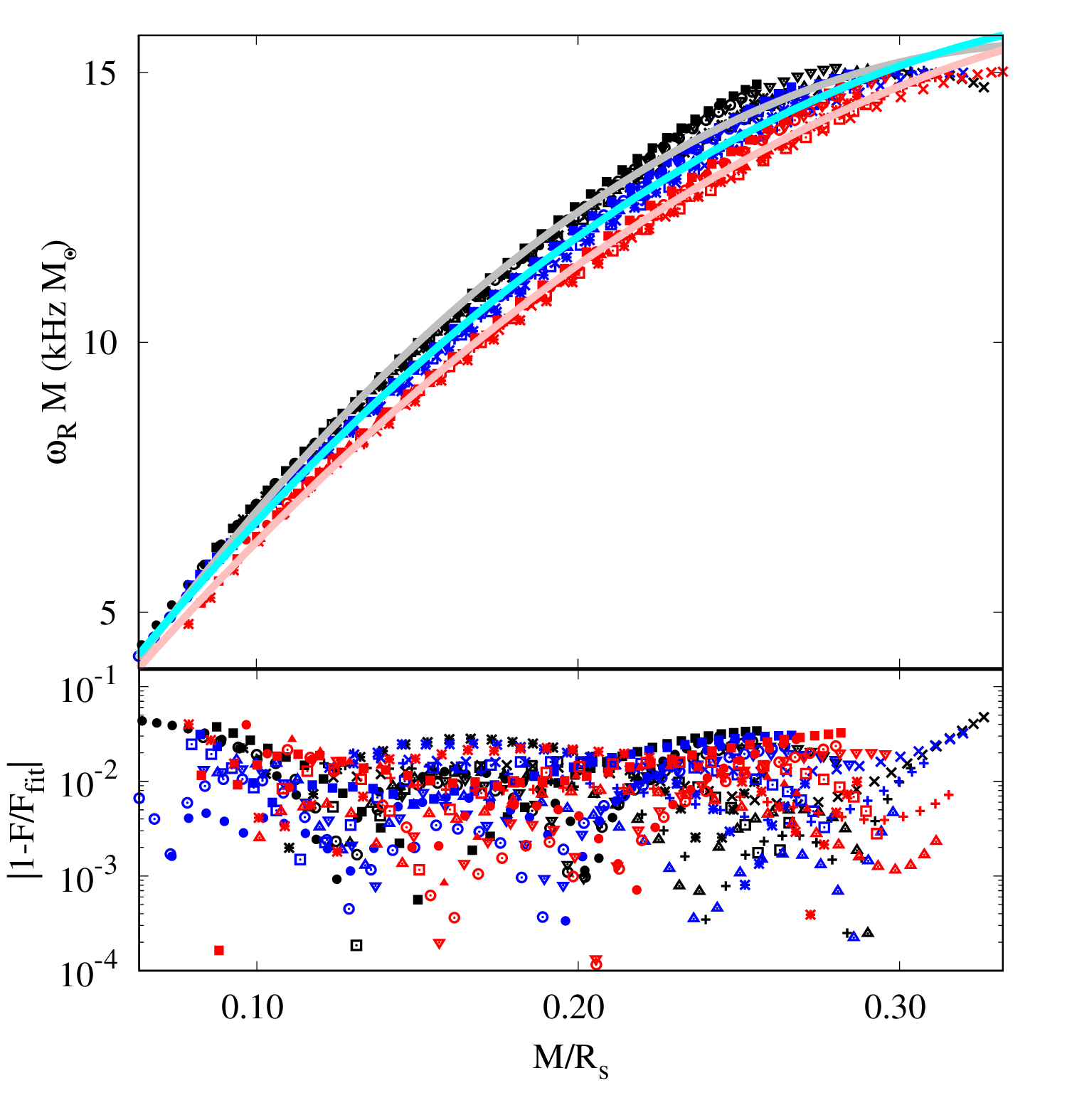}%
	\includegraphics[width=0.49\textwidth,angle=-90]{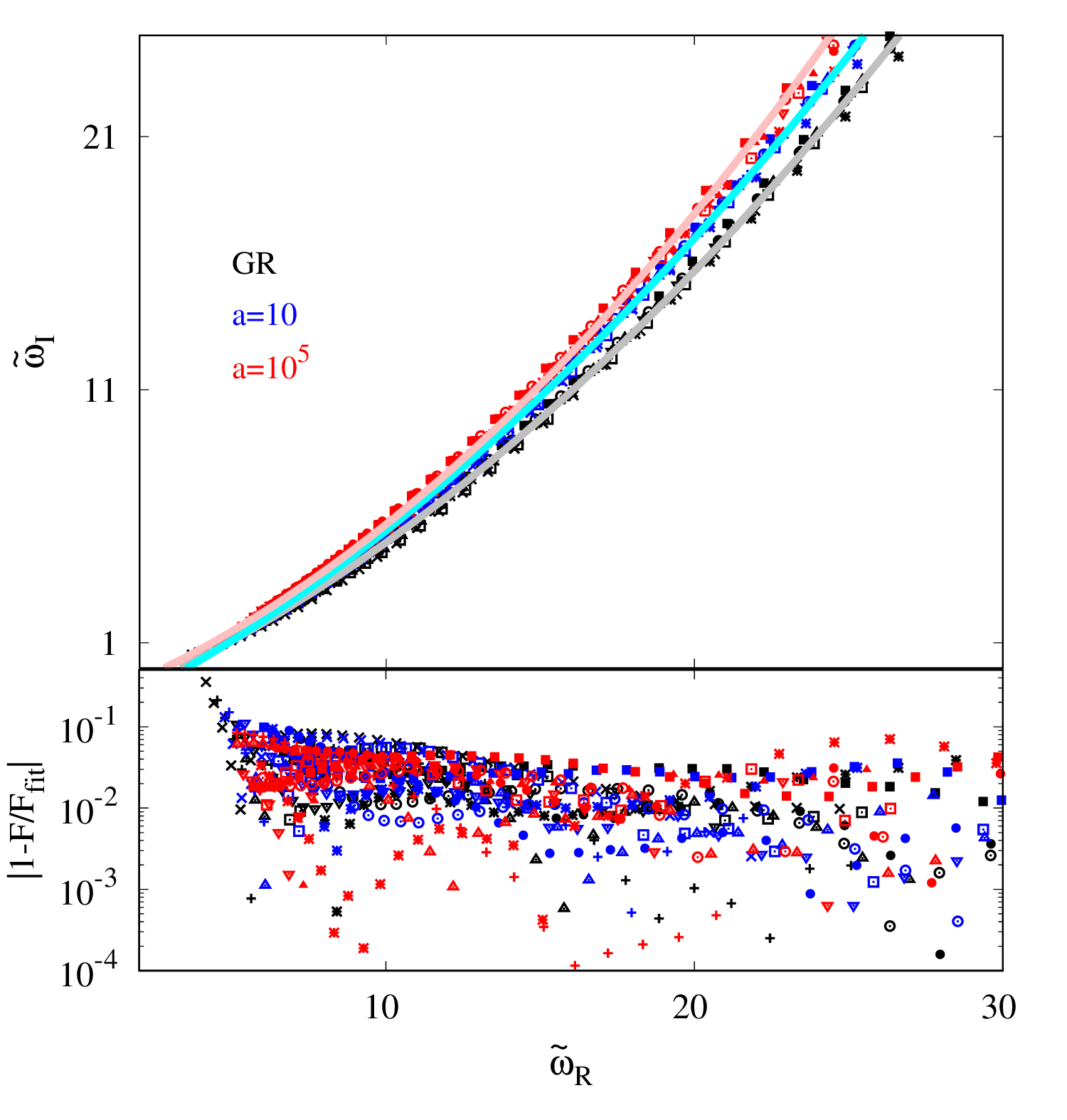}
	\caption{(left) Frequency $\omega_R$ scaled with the mass 
$M$ of the star ($\rm kHz \cdot M_{\odot}$) versus compactness $M/R_s$.  (right) Imaginary part $\tilde{\omega}_I$ versus 
real part $\tilde{\omega}_R$ of the QNM mode, both scaled with the square root
of the central pressure,
for GR (red), $a=10$ (blue) and $a=10^5$ (red). The solid lines correspond 
to the fits (\ref{fit_freq2}) and (\ref{fit_pc}) (in grey for GR, in cyan for $a=10$ 
and in pink for $a=10^5$).}
\label{l2_omR_omI_pc}
\end{figure}

An alternative universal relation for the frequency is shown in Fig.~\ref{l2_omR_omI_pc} (left), where the frequency is
scaled with the total mass of the star instead of the radius. The corresponding phenomenological relation is
\begin{eqnarray}
\omega_R[\rm kHz]\cdot M [\rm M_{\odot}] = 
\begin{cases}
(-1.455\pm 0.108) + (97.03\pm 1.15) \frac{M}{R_s} - (138.5\pm 2.9) \left( \frac{M}{R_s} \right) ^2 &\textnormal{GR}\\
(-0.725\pm 0.085) + (84.62\pm 0.96) \frac{M}{R_s} - (106.0\pm 2.5) \left( \frac{M}{R_s} \right) ^2 &a=10\\
(-0.660\pm 0.108) + (78.73\pm 1.11) \frac{M}{R_s} - (91.3\pm 2.7) \left( \frac{M}{R_s} \right) ^2 &a=10^5
\end{cases}
\label{fit_freq2}
\end{eqnarray}
which possesses a good EOS universality, comparable to the relation (\ref{fit_freq}).

Finally, in Fig.~\ref{l2_omR_omI_pc} (right) we show another interesting 
universal relation, which holds between the real and the imaginary part 
of the QNMs, when both parts are scaled with respect 
to the central pressure of the models
\cite{BlazquezSalcedo:2012pd,Blazquez-Salcedo:2013jka,Blazquez-Salcedo:2015ets}.
Thus defining the dimensionless scaled quantity 
$\tilde{\omega} = \omega/\sqrt{\bar{p}_0}$,
where $\bar{p}_0$ is the central pressure, 
one finds the phenomenological relation
\begin{eqnarray}
\tilde{\omega}_I = 
\begin{cases}
(-1.821\pm 0.067) + (0.482\pm 0.010) \tilde{\omega}_R + (0.0197\pm 0.0003) \tilde{\omega}_R^2 &\textnormal{GR}\\
(-2.278\pm 0.060) + (0.578\pm 0.007) \tilde{\omega}_R + (0.0192\pm 0.0002) \tilde{\omega}_R^2 &a=10\\
(-1.502\pm 0.114) + (0.465\pm 0.016) \tilde{\omega}_R + (0.0254\pm 0.0004) \tilde{\omega}_R^2 &a=10^5
\end{cases}
\label{fit_pc}
\end{eqnarray}
Also this relation shows a rather good EOS universality 
with deviations up to about 10\%.

\section{Conclusions}

We considered axial quasi-normal modes of neutron stars in $f(R)$ gravity 
with various realistic equations of state describing pure nuclear matter, 
hyperon matter, and a mixture of quark and nuclear matter. 
The equation for the axial quasi-normal modes of neutron stars 
in $f(R)$ gravity were derived, and the QNM frequencies 
were calculated numerically. 
Most studies of the spectrum and the universal (EOS independent) 
relations were performed on the basis of the results obtained 
by solving the boundary value problem for the time-independent equation. 
In addition, in order to verify the results 
and to understand the problem in more detail, 
a time evolution of the time-dependent perturbation equation
was performed.

With the time evolution method we studied the wave profiles 
and their deviations from the pure GR ones for several values 
of the free parameter of the theory for neutron star models with a fixed mass. 
From the obtained signal we extracted the oscillation frequency 
and the damping times. The results were compared to the ones 
from the boundary value problem, yielding very good agreement 
between both methods.   

We employed the time-independent perturbation equation to obtain 
the frequencies and the damping times for sequences of models in GR 
and $f(R)$ gravity and analyzed the results in order to reveal the effect 
of the modification of the theory and the value of its free parameter $a$
on the oscillation frequencies and the damping times. 
The effect on the frequencies found for the largest 
considered value of $a=10^5$ 
is a decrease of the frequency of about 8 to 10\% compared to GR, 
and it is more or less constant for all considered masses. 
The damping times decrease as well {when increasing the parameter $a$}. 
Interestingly, the damping times are very close for all values of $a$ 
for small neutron star masses, and the deviation increases 
with the mass of the models. 

We further showed that the previously found universal relations in GR 
remain to a large extent also EOS independent for $f(R)$ gravity.
However, there are qualitative differences 
that increase with increasing parameter $a$. 
Expectedly, the maximal deviation 
arises for the maximal value of $a$ considered, $a=10^5$, 
and the relations tend to the GR ones when $a$ is decreased.
Clearly, one should next also study the polar modes of the
models in $f(R)$ gravity, as well as the fluid modes.
A first step in the latter direction has already been taken
\cite{Staykov:2015cfa}.

\section*{Acknowledgements}
We would like to thank K. Kokkotas for valuable comments.
JLBS and JK would like to acknowledge support by the 
DFG Research Training Group 1620 {\sl Models of Gravity} 
and the COST Action CA16104.
KS, SY, and DD would like to thank for support 
by the COST Action CA16214. DD would like to thank the European Social Fund, the Ministry of Science, Research and the Arts Baden-W\"urttemberg for the support. DD is indebted to the Baden-W\"urttemberg Stiftung for the financial support of this
research project by the Eliteprogramme for Postdocs.

\bibliographystyle{unsrt}

\end{document}